\documentclass[12pt]{iopart}

\usepackage[comma,square,numbers,sort&compress]{natbib}
\usepackage{lineno}
\usepackage{xspace}
\usepackage{xcolor,soul}
\usepackage{hyperref}
\usepackage{xparse}
\usepackage{xstring}
\usepackage{euscript}
\usepackage[utf8]{inputenc}
\usepackage[english]{babel}
\usepackage{graphicx}
\usepackage{blindtext}
\usepackage{float}
\usepackage{adjustbox}
\usepackage{enumitem}
\usepackage{amssymb}
\newsavebox{\measurebox}

\usepackage{iopams}  

\newcommand{\Jpsi}         {\ensuremath{\rm{J/\psi}}\xspace}

\newcommand{\pp}           {pp\xspace}

\newcommand{\PbPb}         {\mbox{Pb--Pb}\xspace}

\newcommand{\pPb}          {\mbox{p--Pb}\xspace}
\newcommand{\Pbp}          {\mbox{Pb--p}\xspace}

\newcommand{\nineH}        {$\sqrt{s}~=~0.9$~Te\kern-.1emV\xspace}
\newcommand{\seven}        {$\sqrt{s}~=~7$~Te\kern-.1emV\xspace}
\newcommand{\twoH}         {$\sqrt{s}~=~0.2$~Te\kern-.1emV\xspace}
\newcommand{\twosevensix}  {$\sqrt{s}~=~2.76$~Te\kern-.1emV\xspace}
\newcommand{\five}         {$\sqrt{s}~=~5.02$~Te\kern-.1emV\xspace}
\newcommand{\twosevensixnn}{$\sqrt{s_{\mathrm{NN}}}~=~2.76$~Te\kern-.1emV\xspace}
\newcommand{\fivenn}       {$\sqrt{s_{\mathrm{NN}}}~=~5.02$~Te\kern-.1emV\xspace}

\newcommand{\GeVc}         {Ge\kern-.1emV/$c$\xspace}
\newcommand{\MeVc}         {Me\kern-.1emV/$c$\xspace}
\newcommand{\TeV}          {Te\kern-.1emV\xspace}
\newcommand{\GeV}          {Ge\kern-.1emV\xspace}
\newcommand{\MeV}          {Me\kern-.1emV\xspace}
\newcommand{\GeVmass}      {Ge\kern-.2emV/$c^2$\xspace}
\newcommand{\MeVmass}      {Me\kern-.2emV/$c^2$\xspace}
\newcommand{\lumi}         {\ensuremath{\mathcal{L}}\xspace}

\begin{document}

\title[Vector meson photoproduction in UPCs with FoCal]{Vector meson photoproduction in UPCs with FoCal}

\author{A. Bylinkin$^1$, J. Nystrand$^1$ and
D. Tapia Takaki$^2$}
\address{$^1$ Department of Physics and Technology, University of Bergen,
Bergen, Norway}
\address{$^2$ Department of Physics and Astronomy, University of Kansas, Lawrence, KS, USA}

\ead{alexandr.bylinkin@cern.ch}

\vspace{10pt}

\begin{abstract}
We discuss the physics prospects of photon-induced measurements using the high-granularity FoCal detector to be installed at the ALICE experiment, covering the pseudorapidity interval $3.4 \leq \eta \leq 5.8$. This new detector, scheduled to be in operation from Run 4, will explore the small Bjorken-$x$ physics region in an unprecedented way. In this region the gluon saturation phenomenon is expected to be dominant. Combined with the rest of the ALICE subdetectors, including the zero degree calorimenters, FoCal will serve to reconstruct in a model-independent way the measured photoproduction cross sections for vectors mesons in a wide range of photon-target energies, down to $x$ values of about $7\times 10^{-6}$ and $2\times 10^{-6}$ in ultra-peripheral photon--proton and photon--lead collisions, respectively.  
\end{abstract}

%
\vspace{2pc}
\noindent{\it Keywords}: ultra-peripheral collisions, UPC, gluon saturation, LHC Run 4 upgrades
%
%
%
%

\section{Introduction}
The FoCal detector~\cite{ALICE:2020mso} of ALICE (A Large Ion Collider Experiment)~\cite{ALICE:2008ngc} at CERN's Large Hadron Collider will address fundamental physics questions about the dynamics of partons inside nucleons and nuclei, exploring a unique kinematic region at high energies where the growth of the gluon density is expected to recede. This is a phenomenon known as gluon saturation that is predicted by unitary principles in the theory of Quantum Chromodynamics (QCD), the most compelling theory for describing the strong interactions of quarks and gluons. Gluon saturated matter implies reaching a dynamical equilibrium between the splitting and recombination of gluons. Such a regime can be modelled with the introduction of non-linear QCD evolution equations~\cite{Gribov:1983ivg, Mueller:1989st,Golec-Biernat:1998zce}. It is of fundamental interest to experimentally determine the saturation scale where this phenomenon appears, for different configurations in nucleons and nuclei. 

It is also of great interest to explore whether the existence of gluon saturated protons implies the formation of a new state of matter with universal features present in hadronic matter at high energies~\cite{Gelis:2010nm}. While measurements at the HERA ep collider at the Deutsches Elektronen-Synchrotron (DESY) have provided evidence for a significant rise of the gluon density in protons as a function of energy or Bjorken-$x$~\cite{H1:2015ubc}, the explored energies were not sufficient to reach a consensus opinion on the formation of gluonic saturated matter. 

Early measurements from fixed target experiments, followed by studies at the Relativistic Heavy-Ion Collider (RHIC) in Brookhaven National Laboratory (BNL)~\cite{BRAHMS:2004xry,STAR:2006dgg,PHENIX:2011puq,PHOBOS:2009sau} indicate a reduction of the density of low momentum gluons in heavy nuclei, as compared to the density in individual protons at high energies. This experimental observation has been explained in terms of nuclear gluon shadowing, a multi-re-scattering process present in nuclei that can be linked to gluon saturation in nucleons~\cite{Armesto:2006ph}. The dynamics of the ``initial state" of nuclei before they collide impacts many measurements in nucleus-nucleus collisions and is important for assessing the properties of the Quark Gluon Plasma (QGP), a phase of QCD matter where quarks and gluons are deconfined. For recent reviews see Refs.~\cite{Busza:2018rrf,Morreale:2021pnn}.

The future Electron Ion Collider (EIC)~\cite{Accardi:2012qut} planned to be built at BNL will focus on addressing several fundamental questions in nuclear physics~\cite{AbdulKhalek:2021gbh}, including the study of gluon saturated matter. Being a dedicated QCD accelerator, it promises detailed studies in a wide variety of reactions, colliding species and energies. At the same time, several unique measurements in small Bjorken-$x$ physics can only be carried out using the high energy beams provided by the LHC. Since the end of the HERA collider, the highest photon energies are being explored using ultra-peripheral heavy ion collisions (UPC) at the LHC. The protons and heavy ions accelerated at the LHC carry themselves electromagnetic fields which can be treated as a flux of quasi-real photons. Using UPCs, the LHC can explore $\gamma \mathrm{p}$ center of mass energies ($W_{\gamma \mathrm{p}}$) up to several \TeV. The center of mass energies for $\gamma \rm{A}$ and two-photon interactions go up to $\sim~$700 \GeV/nucleon and up to $\sim~$150 GeV, respectively. Altogether, such high energy collisions open the possibility to study QCD physics and carry out searches for physics beyond the standard model in an unexplored kinematic region~\cite{Baltz:2007kq,Contreras:2015dqa,Klein:2017nqo,Klein:2019qfb,Klein:2020fmr,Schafer:2020bnm,Frankfurt:2022jns}.

The ALICE collaboration has recently completed a major detector and system upgrade. The new data taking period has started this year with Run 3 (2022--2025) and will be followed by Run 4 (2029--2032). The current ALICE plans are to collect data from \PbPb\ collisions at $\sqrt{s_{\mathrm{NN}}}=5.36$~\TeV, \pp\ collisions at various collision energies, \pPb\ collisions at $\sqrt{s_{\mathrm{NN}}}=8$~\TeV, and O--O interactions at $\sqrt{s_{\mathrm{NN}}}=6.5$~\TeV~\cite{Citron:2018lsq}. We anticipate an integrated luminosity of about \lumi = 13 nb$^{-1}$ in \PbPb collisions in Runs 3~\&~4, an order of magnitude increase with respect to Runs 1~\&~2. The increase in the number of reconstructed events for ALICE UPC measurements will be much more significant because of the continuous detector read out implemented from Run 3. Despite the excellent performance and high-level productivity of the ALICE UPC program, the use of dedicated triggers in Runs 1~\&~2~ has introduced additional systematic uncertainties to the measurements, and some analyses were not possible to be done due to trigger and bandwidth limitations. The prospects for the UPC measurements at the LHC during Runs 3~\&~4 are discussed in Ref.~\cite{Citron:2018lsq}. Some future measurements have also been highlighted in Refs.~\cite{Klein:2020nvu,Hentschinski:2022xnd}. 

In this context, and to fully exploit the potential of the ALICE experiment in the study of the small-$x$ parton structure of nucleons and nuclei, a high-granularity calorimeter~(FoCal) has been proposed~\cite{ALICE:2020mso}. It is expected to be in operation from Run~4. FoCal is a high-granularity, compact silicon-tungsten (Si+W) sampling electromagnetic calorimeter with longitudinal segmentation backed by a conventional high granularity metal and scintillating hadronic calorimeter. It will be located at a distance of $7$~m from the interaction point, outside the solenoid magnet. The Technical Design Report of FoCal is in preparation and expected to be completed by summer 2023. FoCal has been designed for measuring direct photons at forward rapidity with high precision, as well as jets and $\gamma$-jet and jet-jet events, in \pp and \pPb collisions~\cite{ALICE:2020mso}. 

The study of UPCs is also an integral part of the FoCal physics program.  The study of ultra-peripheral heavy-ion collisions with FoCal will contribute to a more systematic and global study of nucleons and nuclei at high energies in  $\gamma\rm{p}$ and $\gamma\rm{Pb}$ interactions. Although the center of mass energies in these interactions will be lower than in pp and PbPb, photon-induced interactions have a clear advantage in probing the structure of the target. In this document, we first present a brief review of recent UPC vector meson measurements at the LHC. We then provide the expected yields of VM photoproduction in the FoCal acceptance, and discuss some high-profile measurements where FoCal will have a competitive advantage with respect to any other LHC measurement.

\section{Brief review of UPC vector meson photoproduction at the LHC}
The ALICE, ATLAS, CMS and LHCb collaborations have carried out UPC measurements in $\gamma \rm{p}$, $\gamma \rm{A}$, and two-photon interactions. The cross sections for these processes at the LHC are very high. The studies of photoproduced vector mesons have been of particular interest. One important goal of these studies is to determine the nuclear gluon distributions. To leading order perturbative QCD, there is a direct proportionality between the photoproduction cross section of heavy vector mesons and the square of the gluon distribution function~\cite{Ryskin:1992ui}. Various attempts have been made in order to translate the photonuclear production cross section to a gluon distribution, including work towards next-to-leading order calculations of parton distribution functions and generalized parton distributions~\cite{Guzey:2013xba,Flett:2019ept,Flett:2019pux}. 

Exclusive photoproduction of \Jpsi off proton targets has been studied at the LHC by ALICE in \pPb collisions~\cite{ALICE:2014eof,ALICE:2018oyo} and by LHCb in \pp collisions~\cite{LHCb:2013nqs,LHCb:2014acg,LHCb:2018rcm}. For a \Jpsi produced at rapidity $y$ it follows from the kinematics that the photon-proton center of mass energy squared is
\begin{equation}
    \rm{W_{\gamma p}^2 = 2 E_p M_{J/\psi} e^{\pm y}}.
\end{equation}
Here, $E_p$ is the proton energy in the laboratory frame. The $\pm$ in the exponent corresponds to the cases when the photon emitting beam moves in the positive and negative direction, respectively. In \pp collisions, it is equally probable that either beam particle emits the photon, whereas in \pPb collisions the Pb nucleus is the dominant photon source. This two-fold ambiguity in determining $W_{\gamma p}$ is the main reason why \pPb is the preferred configuration to study photoproduction off a proton target. The formula also shows that by going to higher rapidities higher values of $W_{\gamma p}$ are obtained. From the $\gamma \rm{p}$ center of mass energy the corresponding Bjorken-$x$ is determined, $x = \rm{(M_{J/\psi}/W_{\gamma p})^2}$. 

Exclusive \Jpsi production on proton target has been studied at HERA up to center of mass energies of 300~GeV, corresponding to $x \sim 10^{-4}$. 
The ALICE measurement in \pPb collisions has extended the energy range to $W_{\gamma p} = 706$~GeV. 

Results from HERA showed that the photoproduction cross section, $\sigma(\gamma p \rightarrow \Jpsi p)$, increases with center of mass energy following a power law. The result from ALICE, extending down to $x \sim 10^{-5}$, showed no deviation from this behaviour, although the statistical error was large.  

While the LHCb collaboration has done studies for exclusive \Jpsi in pp collisions at $\sqrt{s}=7$~\TeV~\cite{LHCb:2013nqs,LHCb:2014acg} and $\sqrt{s}=13$~\TeV~\cite{LHCb:2018rcm}, their analyses of $\sigma(\gamma \mathrm{p})$ are strongly model dependent because of the ambiguity in the photon direction mentioned above. The photoproduction cross section in a \pp collision can be expressed as 
\begin{equation}
\label{equation2}
\frac{d\sigma}{dy} = n(+y) \sigma(\rm \gamma \rm{p, +y}) + \rm{n(-y}) \sigma(\gamma \rm{p,-y}) \, ,
\end{equation}
where $\rm{n(+y)}$ and $\rm{n(-y)}$ are the photon fluxes for positive and negative rapidity, respectively. For each of the measured $d\sigma/dy$ points, they report two ``solutions" for the photon-proton cross section $\sigma(\rm \gamma p)$: one for the photon emitter and one for the photon target, corresponding to two different $W_{\gamma \mathrm{p}}$. It is well known that the corresponding photon fluxes decrease rapidly as the rapidity increases, thus most of their reported solutions are effectively from low $W_{\gamma \mathrm{p}}$. In their 7 \TeV measurements they have assumed that the proton dissociation background is not $W_{\gamma \mathrm{p}}$ dependent, in contradiction to the measurements performed by H1 and ALICE~\cite{H1:2013okq,ALICE:2014eof}. The 13 \TeV pp results from LHCb incorporated the use of a forward detector called HeRSCheL to mitigate the background from proton dissociation, while the photon direction ambiguity remains~\cite{LHCb:2018rcm}. 

For these reasons, only measurements in UPC \pPb collisions can provide a model independent way for studying the energy dependence of $\sigma({\gamma \mathrm{p}})$. While LHCb and ATLAS have not reported measurements in UPC \pPb collisions, CMS has reported results on exclusive $\rho^{0}$~\cite{CMS:2019awk} and $\Upsilon$~\cite{CMS:2018bbk} photoproduction. FoCal serves this program by extending the exclusive \Jpsi measurements further forward in rapidity than what it is possible using the ALICE muon spectrometer and LHCb, reaching down to $x\sim 7\times 10^{-6}$. 

The measurements of coherent \Jpsi photoproduction cross section in UPC \PbPb collisions by the ALICE, CMS and LHCb collaborations have been found to be compatible with moderate nuclear gluon shadowing, and in good agreement with the central values of the EPS09 nuclear parton
parameterization~\cite{ALICE:2012yye,ALICE:2013wjo,ALICE:2021gpt,CMS:2016itn}. Theoretical effort is underway for incorporating the data in parton distributions~\cite{Flett:2019pux}, albeit some theoretical uncertainties remain~\cite{Eskola:2022vpi}. Contrary to pp collisions, it might be possible to determine the photon direction of the photon emitter and the photon target in \PbPb collisions via the study of the neutron dependence using zero degree calorimeters (ZDCs). Only ALICE, ATLAS and CMS have ZDCs, although no separation of the photon sources has been reported at present. FoCal is special since can reach down to $x\sim 2\times 10^{-6}$ for coherent \Jpsi photoproduction by doing a neutron-dependent analysis.  

\section{Physics measurements in \pPb UPCs with FoCal}
The cross sections and expected yields for exclusive production of heavy VMs, $\rm{p+Pb \rightarrow p+Pb+VM}$, in the dielectron decay channel have been calculated using STARlight\cite{Klein:2016yzr}, assuming the geometrical acceptance of FoCal (both electrons within $3.4 \leq \eta \leq 5.8$). The yields have been calculated for both the p-Pb and Pb-p configurations, assuming an integrated luminosity of $150 \, \rm{nb^{-1}}$ in both cases. The results of these calculations are shown in Table~\ref{table:yieldpPb}. 

Hereafter, for the FoCal projections shown in all figures, we have considered a 60\% detector efficiency for the quarkonia measurements as discussed in the Letter of Intent~\cite{ALICE:2020mso}. A 7\% systematic uncertainty is added for the projected points for exclusive \Jpsi photoproduction, and 15\% for \Jpsi mesons with proton dissociation following the Run 2 measurements performed by ALICE. These numbers include the uncertainty on the luminosity determination. To obtain the uncertainty of the projected numbers, the statistical uncertainties calculated from the expected yields have been added in quadrature with the systematical uncertainties reported in ALICE Run 2 analyses. In addition, we only show data point projections when the yield, before applying efficiency corrections, is at least 80 events per bin in rapidity. The projected points are also smeared randomly according to the Gaussian distribution with the width corresponding to the expected statistics of each measurement. Finally, the FoCal energy reach corresponding to its acceptance is shown as a box on the figures.

\begin{table}[htb]
\small
\caption{\small Cross sections and yields for VMs calculated from STARlight. The top three lines correspond to the case when the proton is moving towards FoCAL, corresponding to the low energy regime ($11 \leq W_{\gamma \rm{p}} \leq 36$~GeV for the \Jpsi). The lower three lines corresponds to the opposite configuration when the Pb-nucleus moves towards FoCAL, corresponding to the high energy regime ($1100 \leq W_{\gamma \rm{p}} \leq 3600$~GeV for the \Jpsi).}
\label{table:yieldpPb}
\begin{center}
\begin{tabular}{llll}
\hline
VM  & $\sigma(\rm p+Pb \rightarrow p+Pb+VM)$  & $\sigma(3.4 \leq \eta_{1,2} \leq 5.8)$ & Yield \\ 
              &                                    & p $\rightarrow$ FoCal                  & p $\rightarrow$ FoCal  \\ \hline 
$\rho^0$       & 35 mb      & 140 nb & 21,000 \\
$\phi$         & 1.7 mb     & 51 nb  & 7,700 \\
\Jpsi          & 98 $\mu$b  & 400 nb & 60,000 \\ 
$\psi(2S)$     & 16 $\mu$b  & 8.9 nb & 1,300  \\ 
$\Upsilon(1S)$ & 220 nb     & 0.38 nb & 60 \\ \hline 
               &            & Pb $\rightarrow$ FoCal & Pb $\rightarrow$ FoCal \\ \hline
$\rho^0$       & 35 mb      & 17 nb & 2,600 \\
$\phi$         & 1.7 mb     & 5.3 nb & 800 \\
\Jpsi       & 98 $\mu$b  & 36 nb  & 5,400 \\      
$\psi(2S)$     & 16 $\mu$b  & 0.53 nb & 80 \\
$\Upsilon(1S)$ & 220 nb     & 0.67 pb & $\sim$ 0 \\ \hline
\end{tabular}
\end{center}
\end{table}

\begin{figure}[thb!]
  \centering
\includegraphics[width=\textwidth]{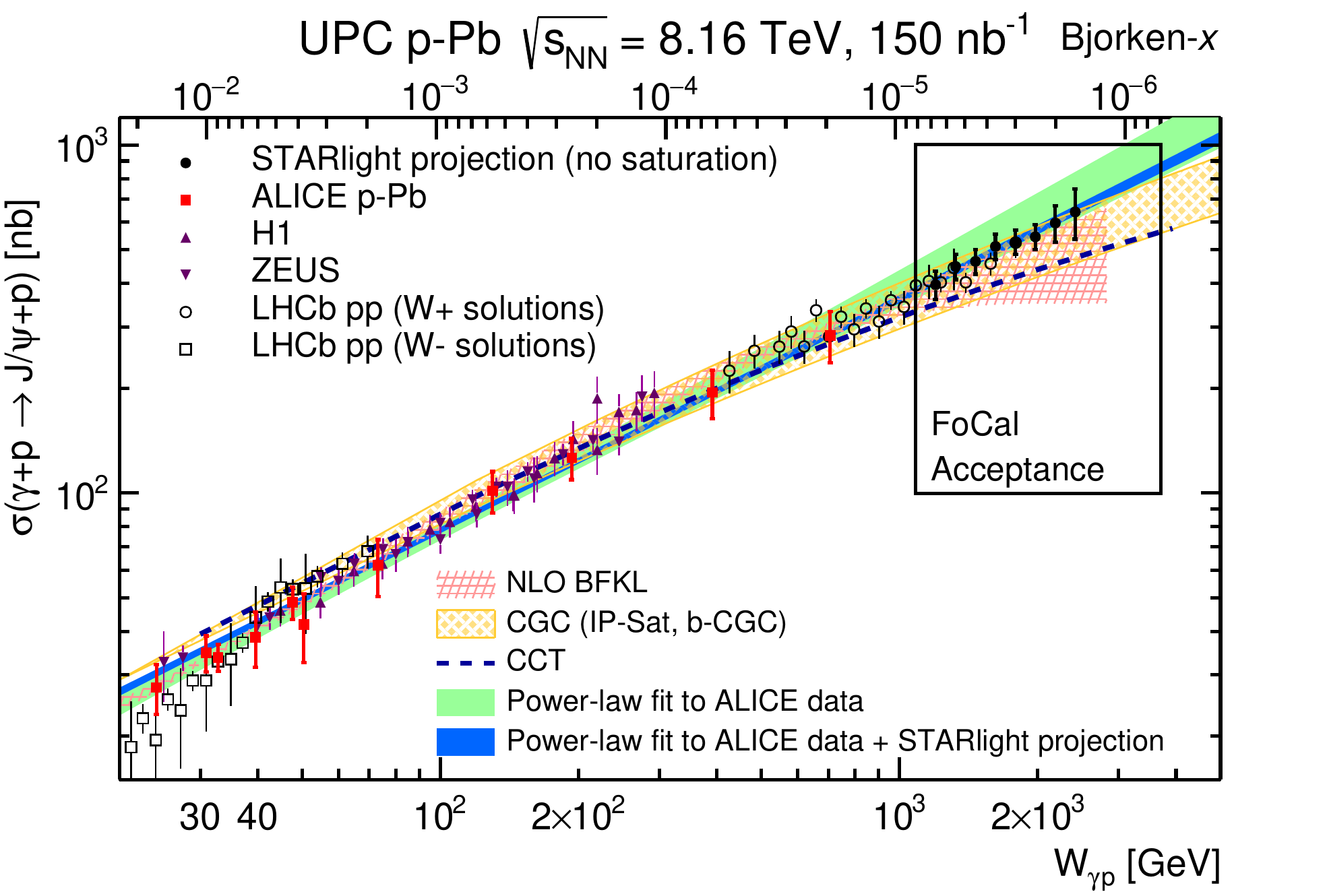}
\caption{\protect\label{fig:pPb1} 
       ALICE data~\cite{Acharya:2018jua} (full red squares) on exclusive photoproduction of \Jpsi off protons as a function of $W_{\gamma \mathrm{p}}$, obtained in \pPb UPCs at $\sqrt{s_{\mathrm{NN}}} = 5.02$ \TeV, compared to a power-law fit (green band), to the data from HERA~\cite{ZEUS:2002wfj, H1:2013okq} (full violet triangles), to the STARlight projection (full black circles) in the FoCal acceptance, and to theoretical models (see text). The LHCb solutions (open circles and squares) for pp collisions at $\sqrt{s} = 7$~\TeV and $13$~\TeV are also shown~\cite{LHCb:2013nqs,LHCb:2014acg,LHCb:2018rcm}. The uncertainties of the ALICE data are the quadratic sum of the statistical and systematic uncertainties. The uncertainties of the STARlight prediction are the quadratic sum of the statistical uncertainty of the expected number of events (Table~\ref{table:yieldpPb}) multiplied by the reconstruction efficiency and the systematic uncertainty taken at the level of ALICE Run 2 measurements~\cite{Acharya:2018jua}. The FoCal energy reach corresponding to its acceptance is illustrated with a box.}
\end{figure}

\begin{figure}[thb!]
  \centering
\includegraphics[width=\textwidth]{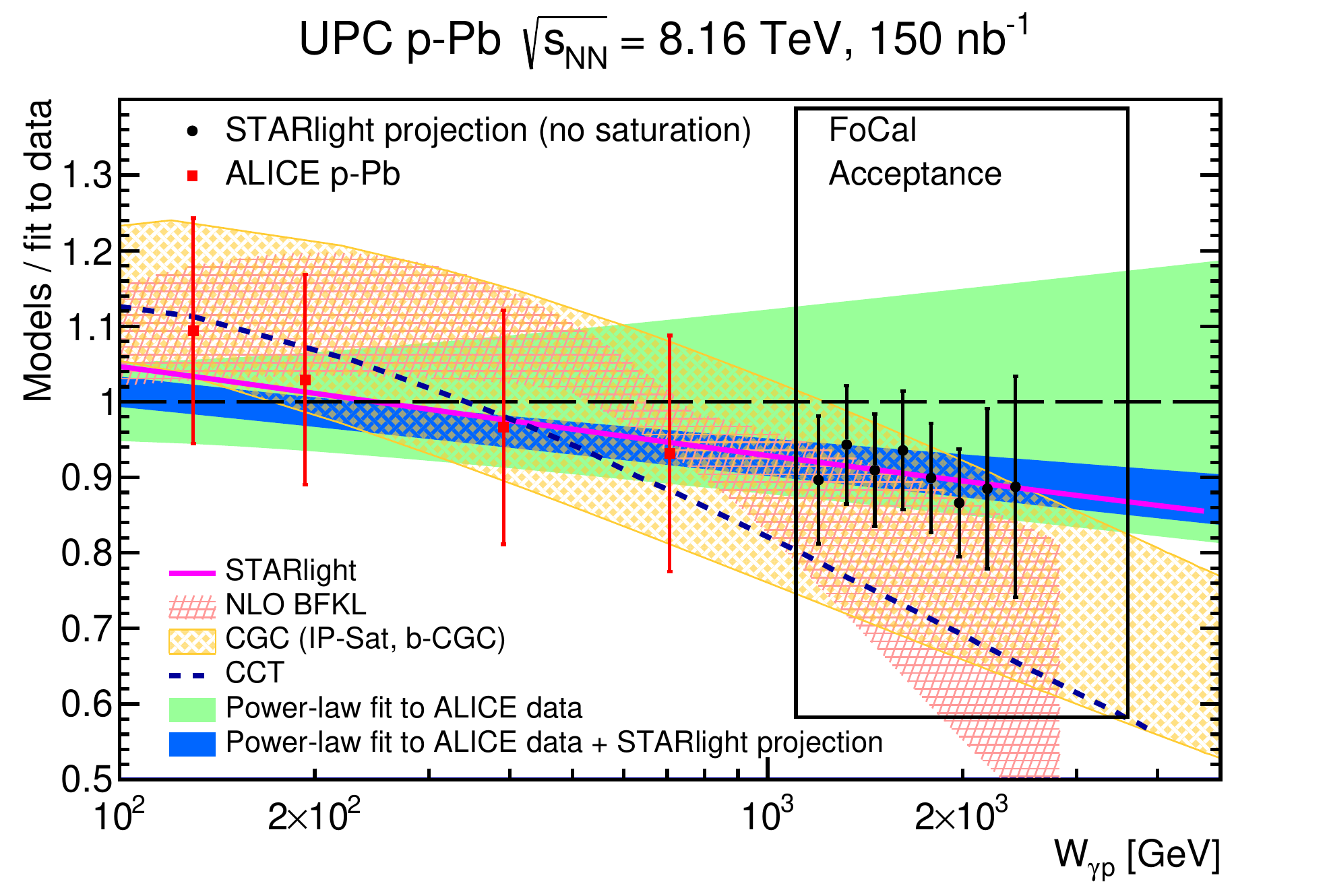}
\caption{\protect\label{fig:pPb2} 
       Ratio of the ALICE data, STARlight projection and theoretical models shown in Fig.~\ref{fig:pPb1} to the power-law fit through the ALICE data points. The green band shows the uncertainty of the power-law fit to the existing ALICE data while the blue band shows the reduction of the fit uncertainty at high energy when the ALICE data is combined with the STARlight projected points. The FoCal energy reach corresponding to its acceptance is illustrated with a box.}
\end{figure}

Figure~\ref{fig:pPb1} shows the energy dependence of exclusive \Jpsi photoproduction measured by ALICE, LHCb, H1 and ZEUS, and the projected data expected by FoCal. The ALICE measurements have been obtained using ultra-peripheral \pPb collisions where there is no ambiguity in the photon direction, which is not the case for symmetric systems such as \pp collisions. As mentioned above, LHCb has made several assumptions for extracting the energy dependence of exclusive \Jpsi in pp. Observing a deviation of a power-law trend would require the most model-independent calculation possible. For this figure, the projected FoCal data are shown for the case when the Pb travels in the direction of the FoCal detector. The STARlight cross section is a parameterization based on the HERA data, which follows a power-law trend. At high energies, there are three sets of non-linear QCD models, namely, the Hot Spot model (CCT)~\cite{Cepila:2016uku}, the NLO BFKL~\cite{Bautista:2016xnp} and the CGC-based calculations~\cite{Armesto:2014sma}. The data are fitted to the power-law function $N(W_{\gamma p}/W_{0})^{\delta}$ with $W_{0} = 90$~\GeV following the HERA measurements and $N$ and $\delta$ being free parameters. The green band shows the experimental uncertainty obtained from the fit to the existing ALICE data alone. The blue band corresponds to the uncertainty obtained from the power-law fit to the ALICE data combined with the STARlight points projected in the FoCal acceptance. Figure~\ref{fig:pPb2} shows the ratio of the existing ALICE data, the STARlight projection and theoretical models to the power-law fit to the existing ALICE data on exclusive \Jpsi photoproduction cross sections. The incorporation of the FoCal data leads to a significant reduction in the measured uncertainty at high energies, even in the case of no saturation scenario (STARlight projection).

\begin{figure}[thb!]
  \centering
\includegraphics[width=\textwidth]{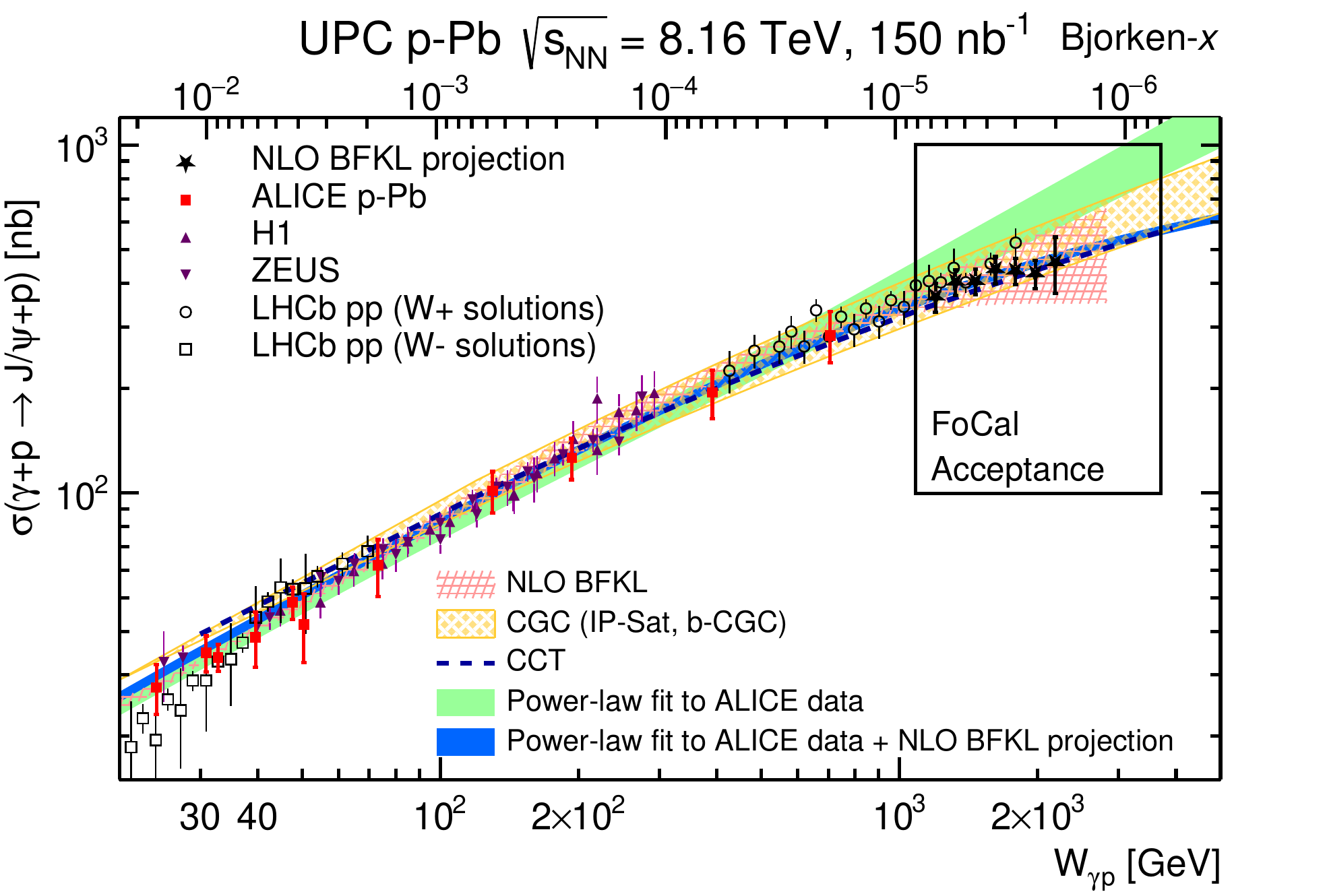}
\caption{\protect\label{fig:pPb1M} 
       ALICE data~\cite{Acharya:2018jua} (full red squares) on exclusive photoproduction of \Jpsi off protons as a function of $W_{\gamma \mathrm{p}}$, obtained in \pPb UPCs at $\sqrt{s_{\mathrm{NN}}} = 5.02$ \TeV, compared to a power-law fit (green band), to the data from HERA~\cite{ZEUS:2002wfj, H1:2013okq} (full purple triangles), to the NLO BFKL projection (full black stars) in the FoCal acceptance, and to theoretical models (see text). LHCb solutions (open circles and squares) for pp collisions at  $\sqrt{s} = 7$~\TeV and $13$~\TeV are also shown~\cite{LHCb:2013nqs,LHCb:2014acg,LHCb:2018rcm}. The uncertainties of the ALICE data are the quadratic sum of the statistical and systematic uncertainties. The uncertainties of the NLO BFKL prediction are the quadratic sum of the statistical uncertainty of the expected number of events (Table~\ref{table:yieldpPb}) multiplied by the reconstruction efficiency and the systematic uncertainty taken at the level of ALICE Run 2 measurements~\cite{Acharya:2018jua}. The FoCal energy reach corresponding to its acceptance is illustrated with a box.}
\end{figure}

\begin{figure}[thb!]
  \centering
\includegraphics[width=\textwidth]{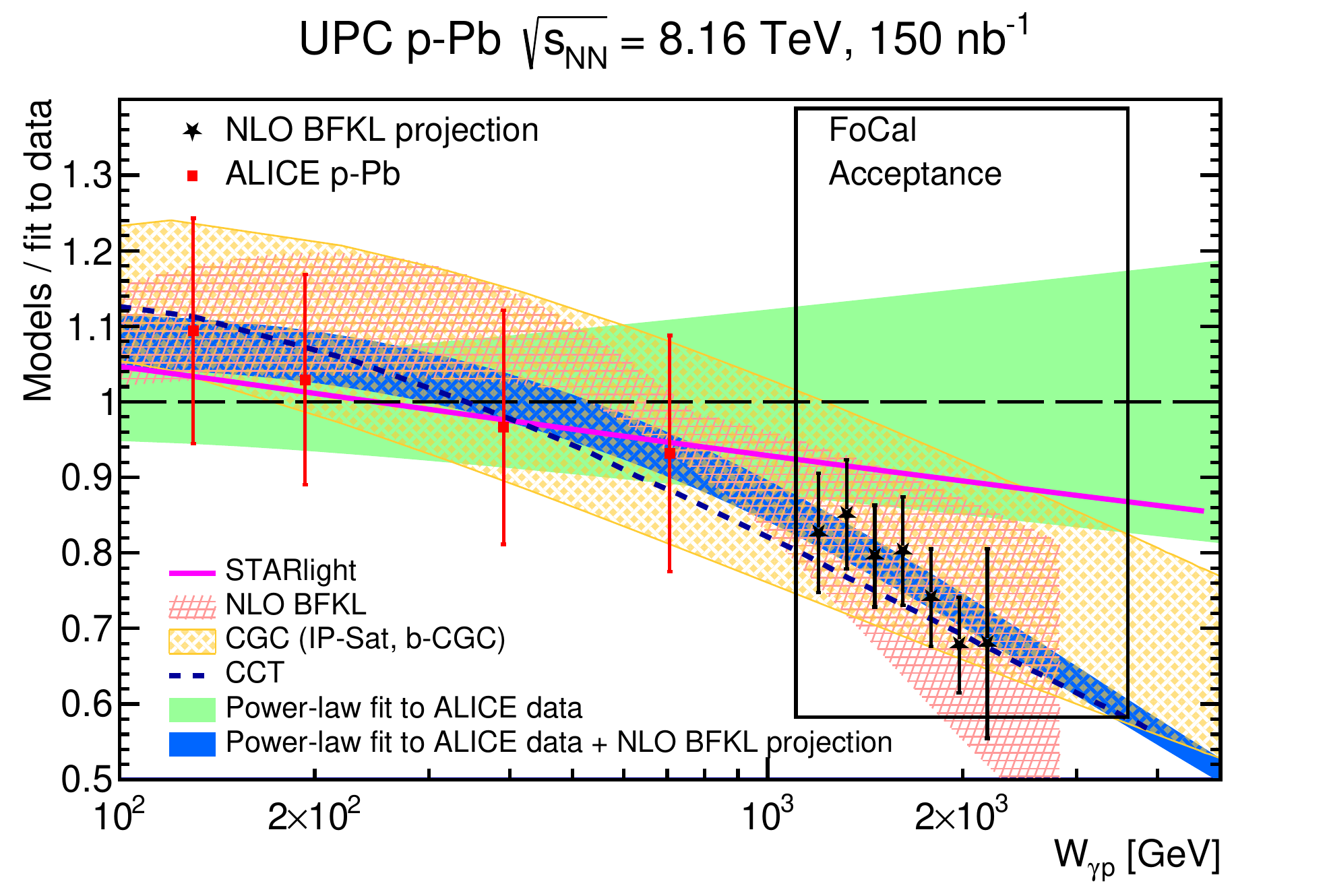}
\caption{\protect\label{fig:pPb2M} 
       Ratio of the ALICE data, NLO BFKL projection and theoretical models shown in Fig.~\ref{fig:pPb1M} to the power-law fit through the ALICE data points. The green band shows the uncertainty of the power-law fit to the existing ALICE data while the blue band shows the reduction of the fit uncertainty at high energy when the ALICE data is combined with the NLO BFKL projected points and fitted to the formula shown in Eq.~\ref{equation3}. The FoCal energy reach corresponding to its acceptance is illustrated with a box.}
\end{figure}

The FoCal detector will provide access to an unexplored kinematic regime at small-$x$ where a different trend in the growth of the cross section might occur. 
For this reason the projection of the \Jpsi photoproduction cross section in the FoCal acceptance was also obtained using the NLO BFKL model~\cite{Bautista:2016xnp}, as it is shown in Figure~\ref{fig:pPb1M}. Since saturation models are expected to deviate from the power-law dependence, the following function is used to fit the existing ALICE data together with the projected data points:
\begin{equation}
\label{equation3}
    \sigma(\gamma p) \approx \frac{N}{\frac{1}{W_{\gamma p}^{\delta}}+ A},
\end{equation}
which has three free parameters: $N$ that gives the overall normalization, $\delta$ that describes the power-law rise of the cross section at lower energies, and $A$ that determines the saturation behavior (broken power-law) at higher $W_{\gamma p}$ values. As one can see, such functional shape provides a good description of both the existing ALICE data and the NLO BFKL projection.
To illustrate it in a more quantitative manner, Figure~\ref{fig:pPb2M} shows the ratio of the NLO BFKL projection to the same power-law fit to the ALICE data (shown in Figure~\ref{fig:pPb1M}). Thus, one can conclude that if saturation occurs as predicted by these models the expected precision of the FoCal measurement on exclusive \Jpsi photoproduction in UPC \pPb collisions would be sufficient to observe the deviation from the power-law trend at high energies. 

\begin{figure}[thb!]
  \centering
\includegraphics[width=\textwidth]{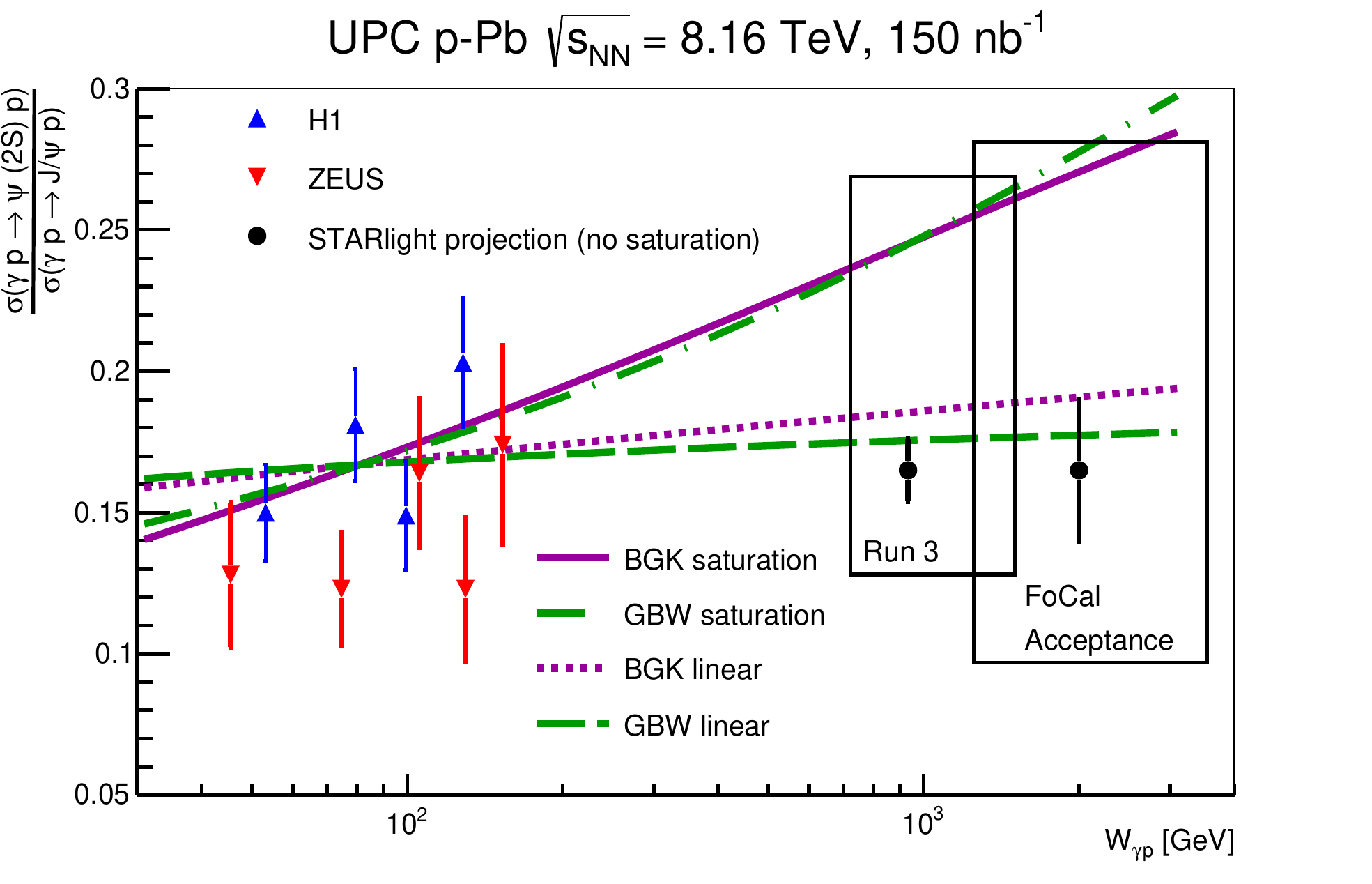}
\caption{\protect\label{fig:pPb3} 
        Energy dependence of the ratio between $\psi(2S)$ and \Jpsi photoproduction cross sections.  The points correspond to the HERA data (full triangles) measured by the H1 and ZEUS collaborations, respectively ~\cite{H1:2002yab, ZEUS:2022sxn}, and to the STARlight projections (full circles). For the STARlight projected points only the statistical uncertainty is shown based on the number of expected events multiplied by the reconstruction efficiency. Theoretical calculations based on BKG and GBW saturation models and their corresponding linearized versions~\cite{Hentschinski:2022xnd}, are also shown. The two boxes illustrate the kinematic regions that can be explored utilizing the ALICE muon detectors during Run 3 and the ALICE FoCal detector during Run 4, respectively.}
\end{figure}

Recent theoretical calculations suggest that the ratio of the cross sections between $\psi(2S)$ and \Jpsi has a larger sensitivity to gluon saturation than the individual cross section themselves~\cite{Hentschinski:2022xnd}. This is partly due to the different wave functions, and the radius dependence of the color dipole for the two mesons. Figure~\ref{fig:pPb3} shows this measured ratio from H1 and ZEUS data~\cite{H1:2002yab, ZEUS:2022sxn}, and the projected STARlight events within the FoCal acceptance. The LHCb collaboration has not reported the measurement for this ratio with systematic uncertainties~\cite{LHCb:2013nqs,LHCb:2014acg,LHCb:2018rcm}, so no LHCb points are shown in Fig.~\ref{fig:pPb3}. The HERA points are compared to two different types of model calculations: the calculations based on the color dipole model (GBW) and on the BGK model, respectively~\cite{Hentschinski:2022xnd}. For these two types of calculations, the linear and the non-linear (saturation) predictions are shown. There is a very visible separation between the linear and the non-linear calculation, making this ratio a very promising  measurement to observe gluon saturation at the LHC. The HERA data at lower energies does not provide any conclusive determination. The projected FoCal points obtained from STARlight agree with the linear-based calculations, which is as expected since it does not include any saturation effects. While this figure does not show the theoretical uncertainty bands displayed on Figure~\ref{fig:pPb2}, the comparison between these two different sets of calculations, the GBW and BGK, provide some confidence that this effect can be seen regardless of which of the two commonly used QCD models is considered. The figure also shows the kinematic region that can be explored utilizing the ALICE muon spectrometer during Run 3. Although there is a sensitivity to observe saturation effects with the Run 3 data alone, a comprehensive science program would require exploring the highest energy points which is only possible with the use of FoCal. Also, there is an added value of having independent measurements using two different detectors at different rapidity intervals in Runs 3\&4. 

\begin{figure}[thb!]
 \vspace*{-0.5cm}
  \centering
\includegraphics[width=\textwidth]{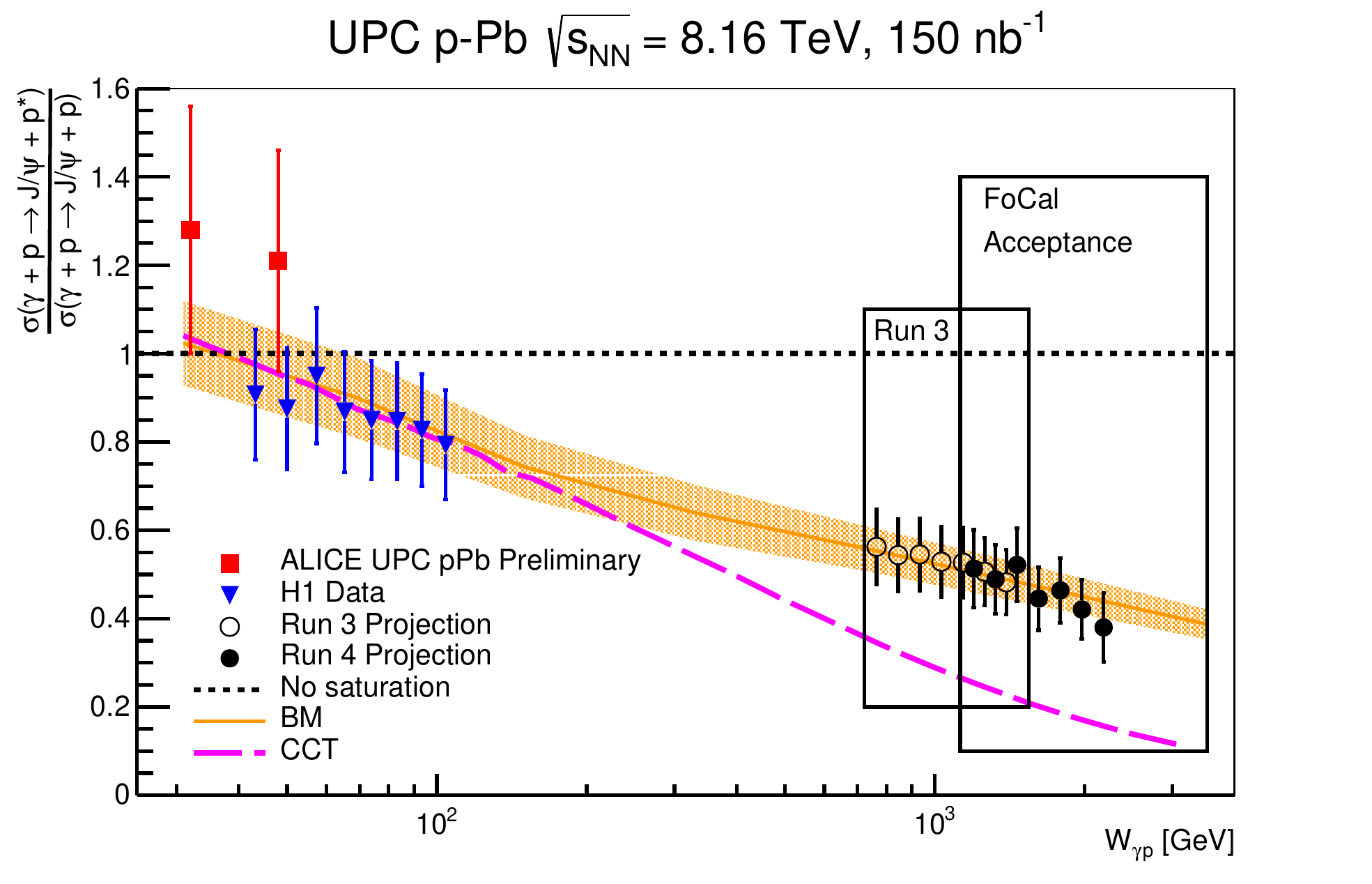}
\caption{\protect\label{fig:pdiss} 
        Energy dependence of the ratio of dissociative to exclusive \Jpsi photoproduction cross sections. The points correspond to the H1 measurement (full triangles), the preliminary ALICE data (full squares) and to the projected data (open and full circles) for the measurements with the ALICE muon detectors during Run 3 and the ALICE FoCal detector during Run 4, respectively. The two boxes correspond to the kinematic regions that can be explored in the FoCal measurements. Theoretical calculations based on the CCT and BM saturation models are also shown.}
\end{figure}

Besides the study of exclusive photoproduction of vector mesons, the dissociative process has received recent interest. According to the Good-Walker formalism~\cite{Good:1960ba, Miettinen:1978jb}, dissociative diffraction probes the variance of the different target configurations, contrary to exclusive processes that only probes the average of such substructures. It has been shown that the momentum transfer distribution of dissociative \Jpsi in proton targets measured by H1~\cite{H1:2013okq} can be described by shape fluctuations (the BM model)~\cite{Mantysaari:2016ykx,Mantysaari:2016jaz}. In addition, the CCT model~\cite{Cepila:2016uku} considers that the proton target configurations measured using the dissociative \Jpsi process can be described by gluonic hot spots, which as the energy increases, their number increases and they start overlapping. When the hot spot overlap is significant, the different configurations look similar and the variance should significantly decrease in the gluon saturation regime. Figure~\ref{fig:pdiss} shows the ratio of dissociative-to-exclusive \Jpsi photoproduction as a function of $W_{\gamma \rm{p}}$, including the H1 data~\cite{H1:2013okq} and ALICE preliminary data, and comparisons to the BM and CCT predictions. The FoCal projected points shown uses the STARlight yield for the exclusive process, and the BM model prediction for the dissociative process. Note that STARlight does not make predictions for dissociative photoproduction. Other more recent predictions for the energy dependence of the proton geometry fluctuations have also been reported~\cite{Kumar:2022aly}. The FoCal data are uniquely positioned to probe the fluctuations of the proton target configurations for gluonic saturated matter. The observation of a significant reduction of the measured cross section as energy increases will provide a clear signature of gluon saturation at high energies.

\begin{figure}[thb!]
 \vspace*{-0.5cm}
  \centering
\includegraphics[width=.9\textwidth]{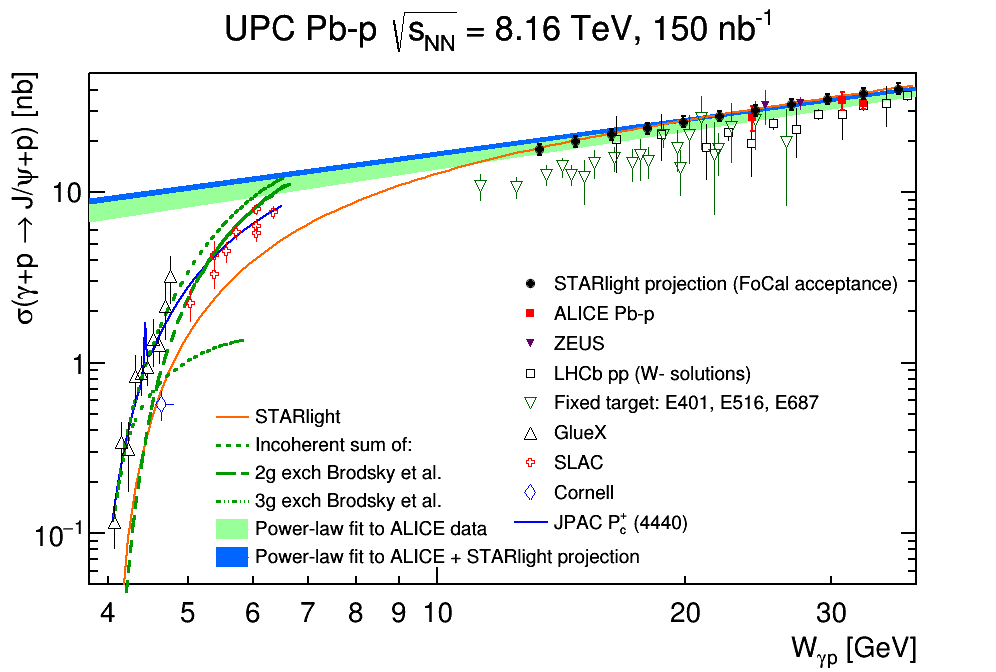}
\includegraphics[width=.9\textwidth]{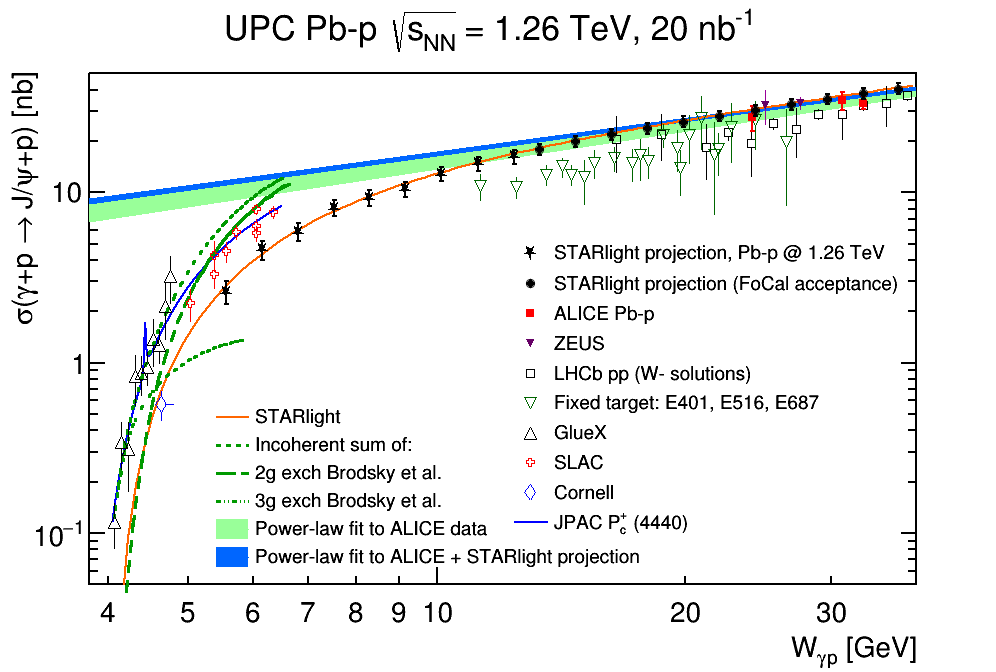}
\caption{\protect\label{fig:lowWpPB} 
         ALICE data~\cite{Acharya:2018jua} (full red squares) on exclusive photoproduction of \Jpsi off protons as a function of $W_{\gamma \mathrm{p}}$, obtained in \Pbp UPCs at $\sqrt{s_{\mathrm{NN}}} = 5.02$ \TeV, compared to a power-law fit (green band), to the data from HERA~\cite{ZEUS:2002wfj} (full blue triangles), to the STARlight projection (full black circles) in the FoCal acceptance, and to theoretical models (see text). LHCb solutions (open squares) for pp collisions at  $\sqrt{s} = 7$~\TeV and $13$~\TeV are also shown together with various measurements performed by fixed target experiments. The lower panel shows in addition the STARlight projection for the \Pbp UPCs at $\sqrt{s_{\mathrm{NN}}} = 1.26$~\TeV (full stars). }
\end{figure}

Using the FoCal detector for the case of ultra-peripheral \Pbp collisions in which the proton travels in its direction might also be of interest. At low energy $W_{\gamma \rm{p}}$, there are several theoretical motivations related to a better description of nucleon mass, quarkonia production and others~\cite{Xie:2021seh}. Figure~\ref{fig:lowWpPB} (top) shows the projected \Jpsi photoproduction cross sections predicted by STARlight, together with the previous LHC, HERA and fixed-target measurements. With FoCal the energies that can be achieved are much lower than those from the current collider experiments, probing as low as $W_{\gamma \rm{p}}$ = 12 GeV, where an onset of near-threshold effects in the \Jpsi photoproduction can already be seen. This measurement will also provide a cross check, with unprecedented experimental precision, of the data reported by the E401 experiment which is found to be significantly below the current theoretical predictions in this kinematic region. Recently, the GlueX experiment at the Thomas Jefferson National Accelerator Facility (JLab) has performed precise measurements of near threshold \Jpsi photoproduction~\cite{GlueX:2019mkq}. Their data are also shown in Fig.~\ref{fig:lowWpPB} as a function of $W_{\gamma \rm{p}}$. There is a significant gap between these low-energy points and the existing LHC data. Before the start of the future EIC, this kinematic region can be explored by having a special \Pbp run at the LHC. We consider a potential special run such that the proton and lead beams would be colliding at $\sqrt{s_{\mathrm{NN}}} = 1.26$~\TeV. It is interesting to point out that with only two days of machine operation, corresponding to a collected luminosity \lumi = 20 {\rm nb$^{-1}$}, the number of recorded events should be sufficient to precisely map out this kinematic region, see Figure~\ref{fig:lowWpPB} (bottom). In this center of mass energy range, the photon energy is low leading to very large fluxes from the lead ion. This explains the large yields even for modest integrated luminosities in this configuration.

\section{Physics measurements in \PbPb UPCs with FoCal} 
Exclusive photonuclear production of VMs has been studied by several experiments at the LHC~\cite{ALICE:2012yye,Abbas:2013oua,Adam:2015gsa,Adam:2015sia}. The cross sections and expected yields for exclusive production of heavy VMs in heavy-ion collisions, $\rm{Pb+Pb \rightarrow Pb+Pb+V}$, in the dielectron decay channel have been calculated from STARlight\cite{Klein:2016yzr}, assuming the geometrical acceptance of FoCal (both electrons within $3.4 \leq \eta \leq 5.8$). The yields have been calculated assuming an integrated luminosity \lumi = $7.0 \, {\rm nb^{-1}}$ for LHC Run 4. The results are shown in Table~\ref{table:yieldPbPb}. The expected statistics for the \Jpsi is very large, and it should thus be possible to map out the detailed shape of the rapidity distribution with high precision over the range $3.4 < y < 5.8$. 

\begin{figure}[thb!]
 \vspace*{-0.5cm}
\includegraphics[width=.9\textwidth]{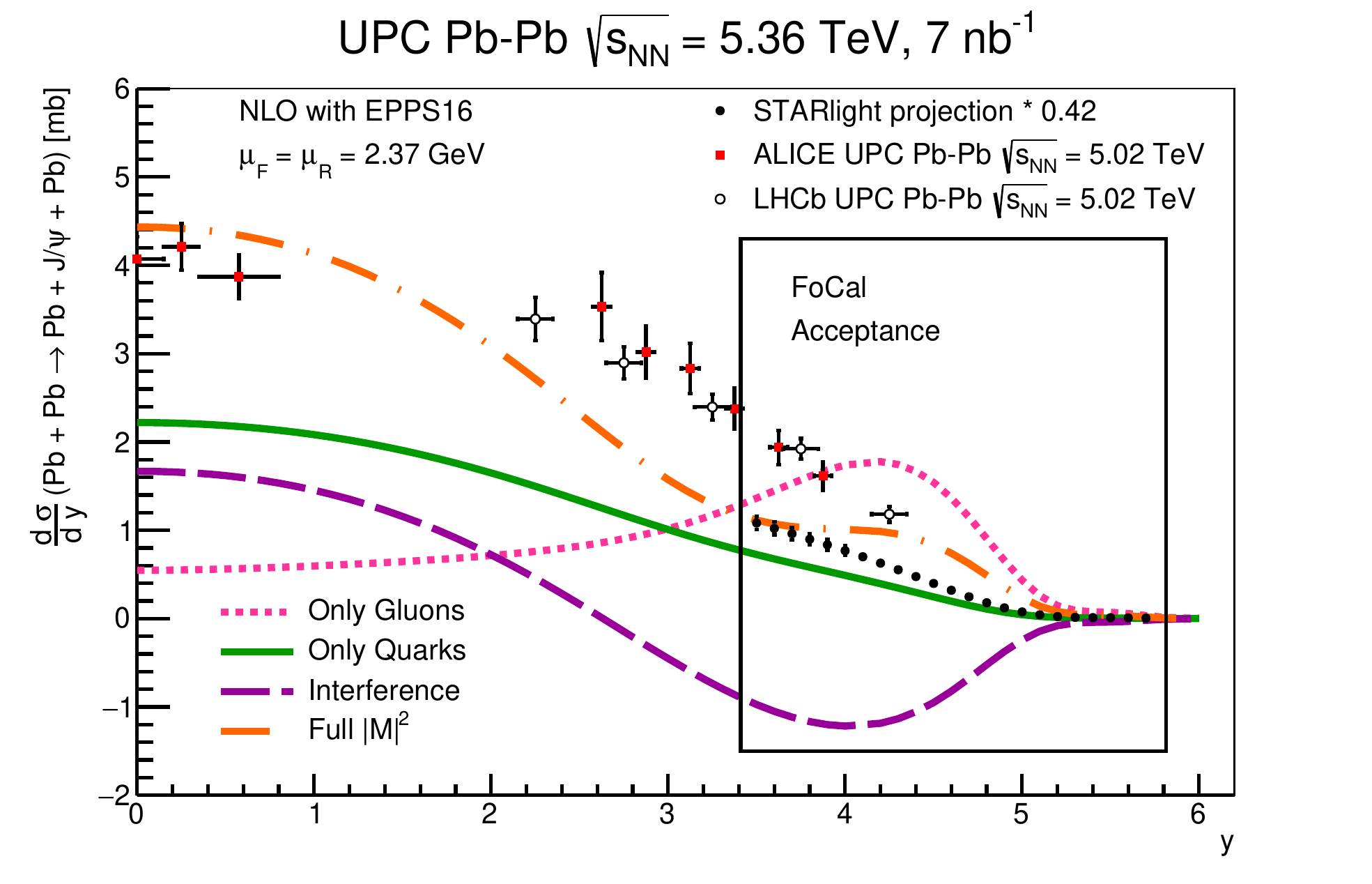}
\includegraphics[width=.9\textwidth]{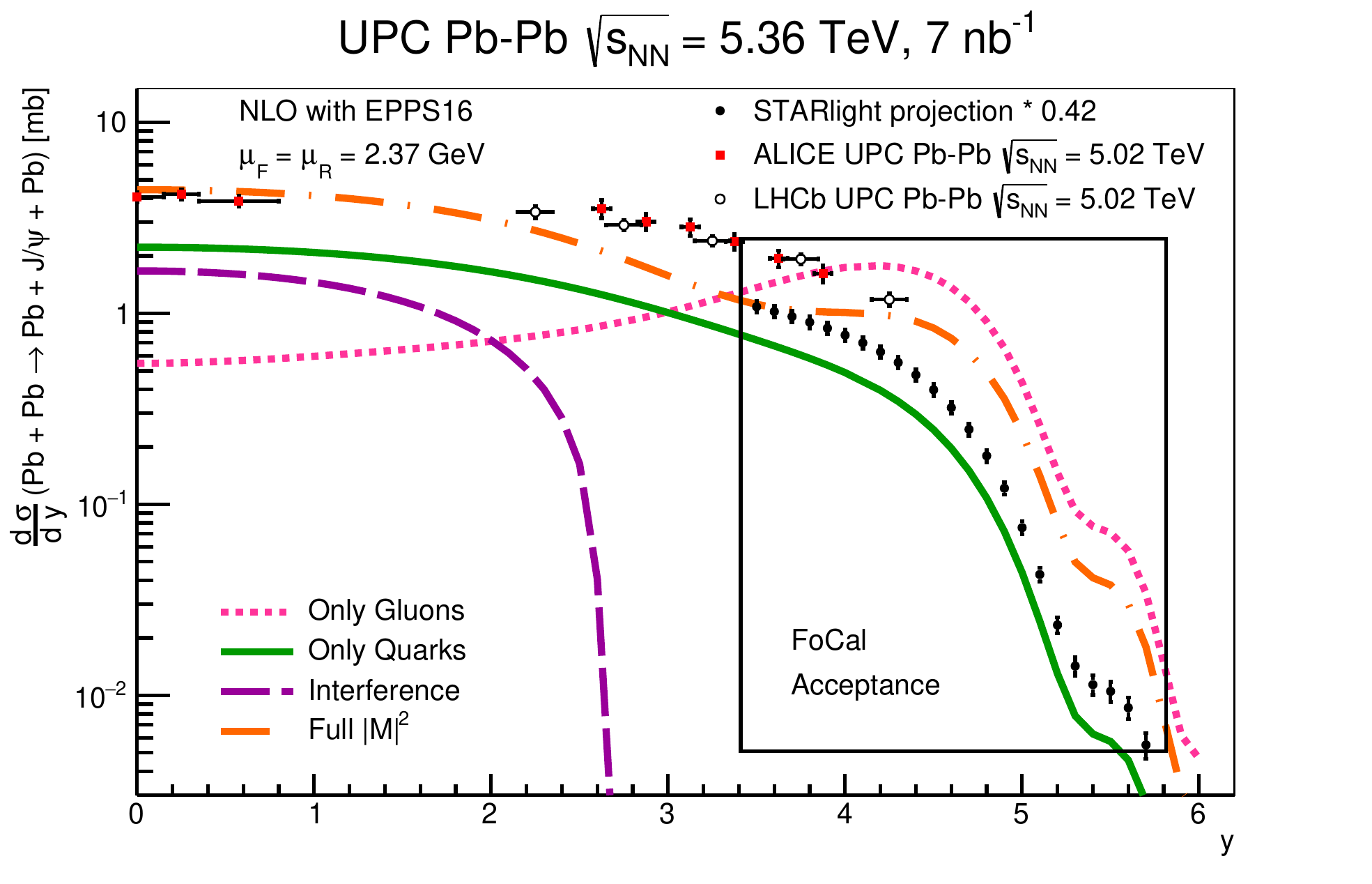}
\caption{\protect\label{fig:PbPb1} 
         ALICE data~\cite{Acharya:2018jua} (full red squares) for the rapidity dependence of the measured cross section for coherent \Jpsi photoproduction in ultra-peripheral \PbPb collisions compared to the recent theoretical calculation (see text)~\cite{Flett:2019pux}, to the LHCb data (open circles) and to the STARlight projection (full black circles) in the FoCal acceptance. The box corresponds to the acceptance of the FoCal detector. The bottom figure shows the same data in the logarithmic scale, where the contribution from the form factor is more visible in the most forward region.}
\end{figure}

A recent study, based on a next-to-leading order calculation, has found that there is a strong contribution to the cross section not just from gluons but also from quarks~\cite{Flett:2019pux}. Figure~\ref{fig:PbPb1} shows the prediction from this model (NLO) compared with the existing ALICE and LHCb data. The figure also includes the STARlight projection into the FoCal acceptance, scaled by a factor 0.42 as documented in the recent CERN Yellow Report~\cite{Citron:2018lsq}. The latter is included since STARlight, which does not include gluon shadowing, is known to overpredict the measured cross section at midrapidity. 
As can be seen in the figure, the interference between the quark and gluon contributions is largest in the region covered by the FoCal acceptance. 

Both STARlight and the NLO calculations show a bump in the cross section at very forward rapidities ($y > 5.3$), see Figure~\ref{fig:PbPb1} (bottom). This bump comes from the second diffractive peak in the nuclear form factor, which cuts off the cross section for very low photon energies. This means that in this region the \Jpsi transverse momentum spectrum will be modified, leading to a larger $<p_T>$ than for the bulk of the events at lower rapidities. 

\begin{table}[htb]
\small
\caption{\small Cross sections and yields for VMs calculated from STARlight assuming an integrated luminosity \lumi = 7.0 {\rm nb}$^{-1}$.}  
\label{table:yieldPbPb}
\begin{center}
\begin{tabular}{llll}
\hline
VM  & $\sigma(\rm Pb+Pb \rightarrow Pb+Pb+VM)$ & $\sigma(3.4 \leq \eta_{1,2} \leq 5.8)$ & Yield \\ \hline 
$\rho^0$       & 5.0 b       & 20 $\mu$b  & 140,000 \\ 
$\phi$         & 440 mb      & 10 $\mu$b  &  70,000 \\ 
\Jpsi       & 39  mb      & 53 $\mu$b  & 370,000 \\ 
$\psi(2S)$     & 7.5 mb      & 1.1 $\mu$b & 7,500  \\ 
$\Upsilon(1S)$ & 94  $\mu$b  & 5.0 nb     & 35 \\ \hline 
\end{tabular}
\end{center}
\end{table}

Contrary to analyses based on \pp data at forward rapidity, it has been suggested that in ultra-peripheral \PbPb collisions it is possible to distinguish the photon direction by studying the emission of forward neutrons that accompany the vector meson~\cite{Citron:2018lsq,Guzey:2013jaa}. The method to identify which of the two Pb nucleus is the photon emitter or the photon target relies on the measurement of \Jpsi events in terms of the following nuclear break up configurations: 0n0n (no neutron on both sides of the interaction point), Xn0n or 0nXn (at least one neutron on one side of the interaction point and no neutrons on the opposite side), and XnXn (at least one neutron on both sides of interaction point). The measured cross section can be expressed as
\begin{equation}
\label{equation4}
\frac{d\sigma}{dy} = \frac{d\sigma\rm (0n0n)}{dy} + 2 \frac{d\sigma\rm (0nXn)}{dy} + \frac{d\sigma\rm (XnXn)}{dy}\, ,
\end{equation}
where for coherent \Jpsi is reasonable to consider that the cross sections are the same for the Xn0n and 0nXn configurations. The relative fractions of the neutron break up modes can be computed using STARlight~\cite{Klein:2016yzr} or nOOn~\cite{Broz:2019kpl}. Table~\ref{table:yieldPbPbNeutrondep} presents the predicted STARlight cross sections and the expected yields. 
\begin{table}[htb]
\small
\caption{\small Cross sections and yields for coherent \Jpsi photoproduction as a function of the different break up scenarios, calculated from STARlight assuming an integrated luminosity \lumi = 7.0 {\rm nb}$^{-1}$. The small difference in the \Jpsi yield compared to Table~\ref{table:yieldPbPb} is due to rounding errors in the calculation.}  
\label{table:yieldPbPbNeutrondep}
\begin{center}
\begin{tabular}{llll}
\hline
Neutron configuration  & $\sigma(\rm Pb+Pb \rightarrow \Jpsi + Pb+Pb)$ & $\sigma(3.4 \leq \eta_{1,2} \leq 5.8)$ & Yield \\ \hline 
0n0n       & 28.8  mb      & 47 $\mu$b  & 329,000 \\ 
0nXn + Xn0n      & 7.3  mb      & 5.0 $\mu$b  & 35,000 \\ 
XnXn       & 3.0  mb      & 2.0 $\mu$b  & 14,000 \\ \hline 
\end{tabular}
\end{center}
\end{table}

For each of three differential equations shown on the left hand side of Eq.~\ref{equation4}, one can express them in terms of the total cross section $\rm \sigma(\gamma + Pb)$ and the corresponding photon flux for both rapidity directions using Eq.~\ref{equation2}, thus resulting in three coupled equations. The solution of this linear system results in the determination of $\sigma(\rm \gamma \rm{Pb, +y})$ and $\sigma(\rm \gamma \rm{Pb, -y})$. In practice, there are several ways how to solve the system of coupled equations. As described in~\cite{Citron:2018lsq,Guzey:2013jaa}, at least the 0n0n and 0nXn are needed since the XnXn component only contributes to less than 4\% of the yield. Also, the XnXn sample might suffer from a large background from non-UPC events. It is also possible to extract the cross sections $\sigma(\rm \gamma \rm{Pb, +y})$ and $\sigma(\rm \gamma \rm{Pb, -y})$ by considering all the measurements together using a $\chi^{2}$ fit or employing the singular value decomposition (SVD) method~\cite{Hocker:1995kb}. 

\begin{figure}[thb!]
 \vspace*{-0.5cm}
  \centering
\includegraphics[width=.9\textwidth]{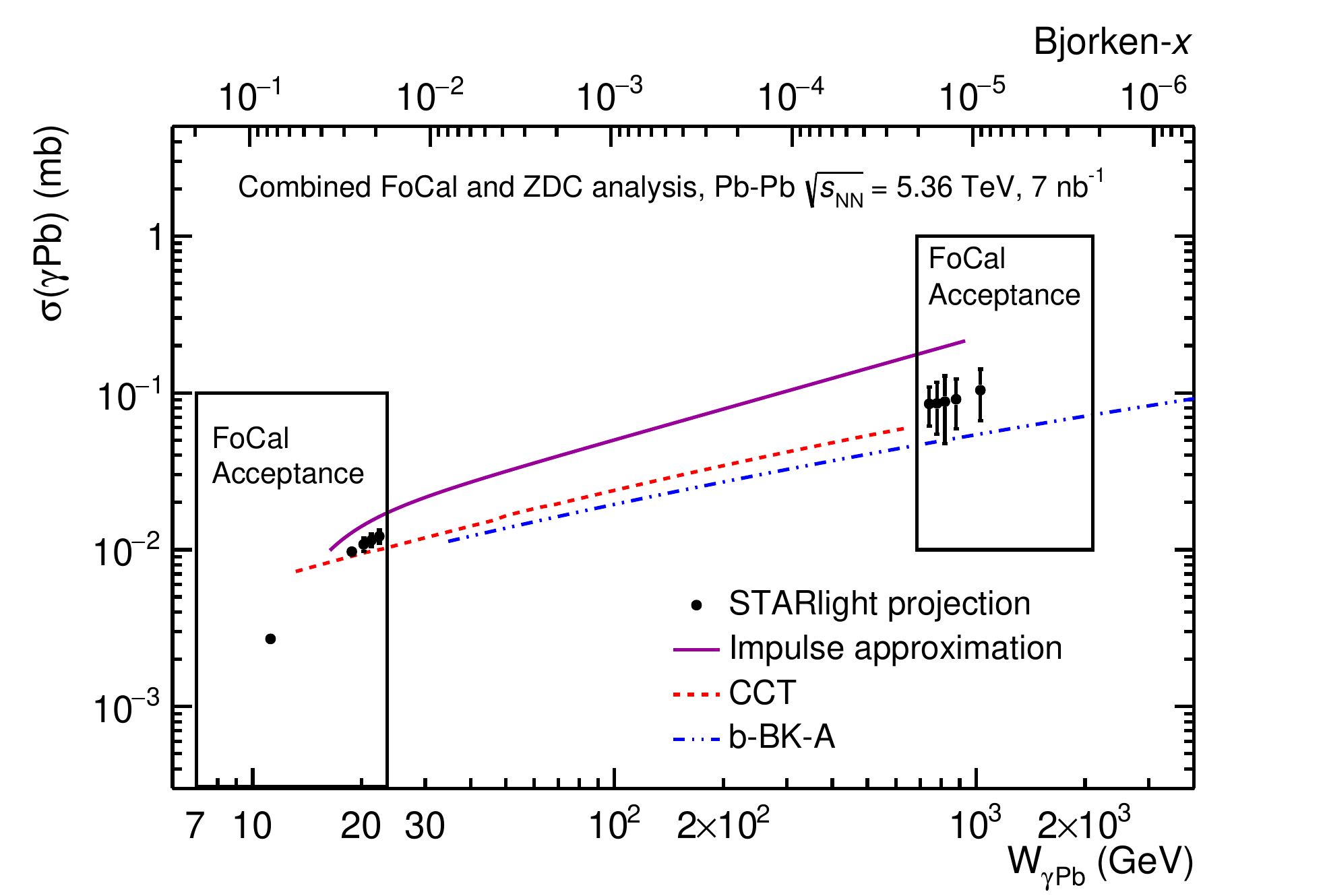}
\caption{\protect\label{fig:PbPb3} 
        STARlight projection (full black circles) for the cross section of coherent \Jpsi photoproduction in ultra-peripheral \PbPb collisions shown as a function of the photon-lead center-of-mass energy $W_{\gamma \rm{Pb}}$. To separate the two photon directions, this analysis will use both the FoCal and ZDC detectors. Predictions from the impulse approximation, CCT and b-BK models are also shown. The two boxes indicate the FoCal detector acceptance for the two explored kinematic regions.}
\end{figure}

For this study, we have considered the SVD method and that the experimental uncertainties of the measurement are about 5\% for 0n0n, about 6\% for 0nXn (or Xn0n) and about 14\% for XnXn. As done above, a 60\% efficiency correction is applied beyond the acceptance effect. We have also required that there should be about 80 \Jpsi events in each of the neutron configurations. This imposes a careful selection of the number of rapidity intervals, thus the $\rm \sigma(\gamma + Pb)$ ranges, that can be studied. This results in different rapidity bin sizes, specially for the lowest and highest energy points. We found that in order to optimize the extraction of the high energy points (a small fraction of the total sample since the photon flux decreases rapidly with rapidity), only five rapidity intervals can be considered, resulting in 10 $\rm \sigma(\gamma + Pb)$ points. Figure~\ref{fig:PbPb3} shows the $\sigma(\gamma \rm{Pb})$ cross section for coherent \Jpsi photoproduction in ultra-peripheral \PbPb collisions as function of $W_{\gamma \rm{Pb}}$, using the SVD method to reconstruct the photon direction. The uncertainty of the measurement results from propagating the experimental uncertainties of the $\rm d{\sigma}/{dy}$s in the various neutron configurations. Clearly a different analysis of the data could also be done so to optimize the low energy points, while here we preferred to illustrate the projected statistics and uncertainties by optimizing the extraction of the high energy points where the effects of non-linear QCD should be dominant. Figure~\ref{fig:PbPb3} also shows the predictions of the impulse approximation, and the CCT and b-BK-A models~\cite{Cepila:2016uku,Cepila:2018zky,Bendova:2018bbb}. The latter models include gluon saturation effects, while STARlight consider Glauber-like corrections not present in the impulse approximation. At present, no experimental data exist for this measurement. Note that LHCb does not have ZDCs installed at present or for Run 4, making ALICE, thanks to the FoCal detector, the only experiment capable of carrying out this analysis, reaching down to $x$ values of about $2\times 10^{-6}$. 

\section{Other possible UPC measurements with FoCal}
There are several additional UPC measurements where FoCal can contribute. Such measurements include the study of low-mass vector mesons such as $\rho^{0}$ and $\phi(1020)$~\cite{Goncalves:2018blz}, dielectron production from two-photon interactions, excited $\rho$ states~\cite{Klusek-Gawenda:2020gwa}, open charm~\cite{Klein:2002wm}, inclusive photonuclear dijets~\cite{ATLAS:2017kwa} and diffractive dijets~\cite{CMS:2022lbi}. The light-by-light scattering process could also be explored~\cite{Klusek-Gawenda:2021upp} in future performance studies. UPC processes have also been suggested as a way to probe the Einstein-Podolsky-Rosen relationship~\cite{Klein:2002gc}. It would be interested to utilize new techniques such as quantum tomography~\cite{Martens:2017cvj} to analyze the UPC data in new ways. Finally, photon-induced interactions might be of future interest in the search for physics beyond the standard model.

\section{Summary}
The high-granularity FoCal calorimeter at ALICE will provide access to an unexplored kinematic region where gluon saturation phenomena should be very visible. We have provided the expected yield of vector meson photoproduction in the dielectron decay channel. The expected statistics will be very large in both ultra-peripheral \pPb and \PbPb collisions. The future measurements of the energy dependence of $\sigma(\gamma \rm{p})$ and $\sigma (\gamma \rm{Pb})$ for the photoproduction of \Jpsi mesons should provide a clear signature of gluon saturation if it is there. Moreover, FoCal will enable measuring the energy dependence of the ratio of exclusive $\psi(2S)$--to-\Jpsi, which is a measurement with high sensitivity. The study of the dissociative production of exclusive \Jpsi is sensitive to quantum fluctuations of the proton constituents, thus also sensitive of gluon saturation. Finally, we have discussed the prospects of studying the low energy points of \Jpsi photoproduction in UPC \pPb collisions, and the impact of having a short special run in \pPb to explore the photoproduction of \Jpsi near threshold. These prospects have been discussed in the context of existing and future measurements at LHC, concluding that FoCal provides a unique physics program for these processes, probing the high-energy limit of QCD in an unprecedented way. 

\section*{Acknowledgments}
We thank colleagues from the ALICE UPC and FoCal groups for interesting discussions, and Martin Hentschinski, Heikki M\"{a}ntysaari and the authors of~\cite{Eskola:2022vpi} for providing their theory predictions. We acknowledge support by the Norwegian Research Council and the U.S. Department of Energy, Office of Science, Nuclear Physics. 

\clearpage
\newcommand{\newblock}{}
\bibliographystyle{utphys}
\bibliography{references}

\end{document}